\documentclass{article}

\usepackage{arxiv}
\usepackage{amsmath}
\usepackage{graphicx}

\title{Manifestation of the anisotropic properties of the molecular J-aggregate shell in optical spectra of plexcitonic nanoparticles}

\author{Alexey~D.~Kondorskiy\thanks{kondorskiy@lebedev.ru}, ~Sergey~S.~Moritaka and Vladimir~S.~Lebedev \\ \\
	P.N. Lebedev Physical Institute of the Russian Academy of Sciences,\\
	Leninskiy prosp. 53, 119991 Moscow, Russia
}

\begin{document}
\maketitle

\begin{abstract}
The theoretical studies of light absorption and scattering spectra of the plexcitonic two-layer triangular nanoprisms and three-layer nanospheres are reported. The optical properties of such metal-organic core--shell and core--double-shell nanostructures were previously explained within the framework of pure isotropic models for describing their outer excitonic shell. In this work, we show that the anisotropy of the excitonic shell permittivity can drastically affect the optical spectra of such hybrid nanostructures. This fact is confirmed by directly comparing our theory with some available experimental data, which cannot be treated using conventional isotropic shell models. We have analyzed the influence of the shell anisotropy on the optical spectra and proposed a type of hybrid nanostructure that seems most convenient for experimental observation of the effects associated with the anisotropy of the excitonic shell. A strong dependence of the anisotropic properties of the J-aggregate shell on the material of the intermediate spacer layer is demonstrated. This allows proposing a new way to effectively control the optical properties of metal-organic nanostructures by selecting the spacer material. Our results extend the understanding of physical effects in optics of plexcitonic nanostructures to more complex systems with the anisotropic and multi-excitonic properties of their molecular aggregate shell.
\end{abstract}

\section{Introduction}

The solution of a number of fundamental and applied problems of nanophotonics is closely related to the study of the optical properties of hybrid organic-inorganic nanostructures and materials. A promising approach to creating novel hybrid materials with unique optical properties is combining various plasmonic nanostructures as an inorganic component and complex molecular or excitonic supramolecular structures as an organic component of composite systems. Typical systems of this kind usually consist of single metallic nanoparticles (or their complexes and arrays) and ordered molecular structures, e.g., J-aggregates of organic dyes~\cite{Wiederrecht2008, Fofang2008, Lebedev2008, Schlather2013, Nan2016, Todisco2018, Kondorskiy2018, Krivenkov2018, Pelton2019}. The optical properties of metallic nanoparticles have been intensively studied in many works \cite{Kreibig1995, Kelly2003, Halas2011, Harris2016, Kond-Lam-Leb_JRLR2018}. The effects of light interaction with such nanoparticles are explained in terms of localized surface plasmon resonances, responsible for great enhancing local electric fields near their surfaces and the light absorption and scattering cross sections. The optical properties of molecular J-aggregates are also well-known \cite{Wurthner2011, Bricks2018, Shapiro_OE2018} and can be successfully described within the framework of the Frenkel exciton model or its generalizations. J-aggregates possess a very narrow optical absorption band, resonance fluorescence with small Stokes shift, anomalously high oscillator strength, and great nonlinear optical susceptibility. 

Over the past two decades, considerable efforts have been devoted to the experimental and theoretical investigations of the structure and optical properties of various two-layer nanoparticles of different sizes, shapes, and compositions, consisting of a metallic core (Ag, Au) and a cyanine dye J-aggregate outer shell. In most of these works, such hybrid particles of spherical~\cite{Kometani2001, Wurtz2003,  Wiederrecht2004, Uwada2007, Lebedev2010, Lekeufack2010, Leb-Medv2012, Leb-Medv2013a, Vujacic2012, Antosiewicz2014} and elongated (spheroids~\cite{Gullen2010, Chen2012, Zengin2013}, rods~\cite{Wurtz2007, Ni2010, Melnikau2016, Thomas2018}) shapes were studied. There are also several papers on two-layer metal-organic nanoplatelets (disks~\cite{ Bellessa2009, Zheng2010, Balci2019}, prisms~\cite{Balci2013, DeLacy2015, Das2017, Lam-Kond-Leb2019} and stars~\cite{Melnikau2013}). Further, these studies were supplemented by a series of works on the optical properties of three-layer nanostructures with a metallic core coated with a double organic shell consisting of an intermediate passive dielectric spacer and an outer layer of cyanine dye molecular J- or H-aggregates. Here it is worthwhile to mention some experimental and theoretical papers on the three-layer nanospheres~\cite{ Yoshida2009a, Yoshida2010, Medv-Leb2010, Leb-Medv2013b, DeLacy2013, Laban2015}, nanorods~\cite{Yoshida2009b, Shapiro2012, Shapiro2015}, nanodumbbells~\cite{Kondorskiy2015}, and nanodisks~\cite{Todisco2015}. The optical properties of such two-layer and three-layer metal-organic nanoparticles are primarily determined by the effects of the near-field electromagnetic coupling of Frenkel excitons in the J-aggregate outer organic shell with localized surface plasmons confined in the metallic core. Quite different regimes of plexcitonic coupling may occur depending on the size and shape of a hybrid nanoparticle and specific optical constants of its metallic and organic subsystems. These are weak, strong, and ultra-strong coupling regimes discussed in several review articles~\cite{Barnes2015, Sukharev2017, Cao2018, Manuel2019} as well as one more coupling regime, which manifests itself in spectral-band replication in the optical spectra of individual plexcitonic nanoparticles~\cite{Lam-Kond-Leb2019} or the nanoparticle dimers~\cite{Kondorskiy2018, Kond-Leb_OE2019}. 

Theoretical description of light absorption and scattering spectra and quantitative analysis of the plasmon-exciton coupling regimes in hybrid metal/J-aggregate and metal/spacer/J-aggregate nanostructures were carried out earlier in almost all cases within the framework of various isotropic models based on the quasistatic approximation or exact analytical and numerical methods, such as the generalized Mie theory, Finite Difference Time Domain (FDTD) method, and some other numerical techniques. Within the framework of the isotropic models, the outer layer of the organic dye is usually considered as an isotropic medium with an effective local dielectric function of the resonance type. In many cases, electrodynamical approaches using the isotropic complex dielectric functions of a metallic core and a J-aggregate shell make it possible to reasonably describe the optical spectra of plexcitonic nanosystems and adequately interpret one or another regime of near-field electromagnetic coupling. Nevertheless, isotropic approaches have some limitations, and their use should be unjustified when the outer organic shell of the system has pronounced anisotropic dielectric properties. 

At the same time, it is well known that structures of nanometric and micrometer sizes made of orientational molecules are generally anisotropic. A vivid example of the dyes capable of forming molecular J-aggregates with pronounced anisotropic dielectric properties is pseudoisocyanine dye (PIC:  1,1'-disulfopropyl-2,2'-cyanine triethylammonium salt). Various morphologies and formations of this and similar molecular aggregates have been the subject of intense studies for many years~\cite{Scherer1984, Misawa1994, Tani2007, Tani2012, Haverkort2014}. In particular, linear dichroism of a spin-coated film dispersed with highly oriented J-aggregates of pseudoisocyanine bromide was investigated to clarify the molecular arrangement of the constituent molecules~\cite{Misawa1994}. Anisotropic behavior of the absorption and fluorescence spectra of fibril-shaped J-aggregates of pseudoisocyanine dyes in thin-film matrices of polyvinyl sulfate was observed in \cite{Tani2007}. Experimental results for oriented J-aggregates of PIC dye demonstrated an essential difference in absorption spectra obtained for polarizations parallel and perpendicular to the J-aggregate orientation axis. Considerable interest in elucidating the role of the anisotropic dielectric properties of the PIC J-aggregate in the formation of the optical properties of hybrid nanosystems is due to the use as an external organic layer in studies of the effects of plexcitonic coupling in metal-organic nanoparticles of various sizes and shapes including concentric spheres~\cite{Uwada2007, Yoshida2009a, Leb-Medv2012, Yoshida2010} and triangular prisms \cite{DeLacy2015, Das2017, Lam-Kond-Leb2019}.

Studies of the effects of anisotropy in the diffraction, absorption, and scattering of light by diverse nanoparticles have been actively carried out for many years. This is due to the important role of these effects in some technological and biological applications~\cite{Joannopoulos2010}. The orientational effects in dipole-dipole interactions between neighboring dye molecules were recently considered using the coupled-dipole model in a spherical shell geometry~\cite{Ru2018a}. A theoretical analysis performed in~\cite{Ru2018a} was carried out along with the study of the influence of the dye molecule concentration in the shell and the investigation of the role of uniformity in its coverage. A thin-shell approximation of the Mie scattering problem for a spherical core--shell and core--double-shell structures with radial anisotropy in the external layer was proposed in \cite{Ru2018b, Tang2021}. Such approximation describes some of the orientational and anisotropic effects arising from the resonant dye-plasmon coupling. In addition, the authors of papers \cite{Ru2019, Tang2020} have developed a model for the evaluation of effective dielectric function for an anisotropic layer of polarizable molecules adsorbed on a metallic surface and for the description of the electromagnetic core-shell interaction in such systems.

Despite this progress, to authors' knowledge, there are no works devoted to the theoretical analysis of the effect of the excitonic shell anisotropy of hybrid metal/J-aggregate and metal/spacer/J-aggregate nanoparticles on their optical properties. The main goal of this work is to elucidate the role of the orientational effects in the molecular arrangements of J-aggregated excitonic shell in the formation of the optical spectra of some plexcitonic core-shell and core--double-shell nanoparticles of various sizes, shapes, and compositions. To this end, we use an approach, which allows us to include the anisotropic dielectric function of a J-aggregate subsystem into the theoretical consideration. The approach takes into account the tensorial nature of the dielectric function of a J-aggregate shell with normal and tangential components. 

This paper is organized as follows. In Sec. 2, we present our theoretical approach and discuss the calculation details, including describing the isotropic local dielectric function of a metallic core and the anisotropic dielectric function of an organic shell. Section 3 is devoted to discussing our results of calculations obtained for the cases of isolated plexcitonic nanospheres and nanoprisms. Particular attention is paid here to comparing the theory with the available experimental data. Finally, in Sec. 4, we give main conclusions of our work.

\begin{figure}[htbp]
\begin{center}
\centering\includegraphics[width=11.0cm]{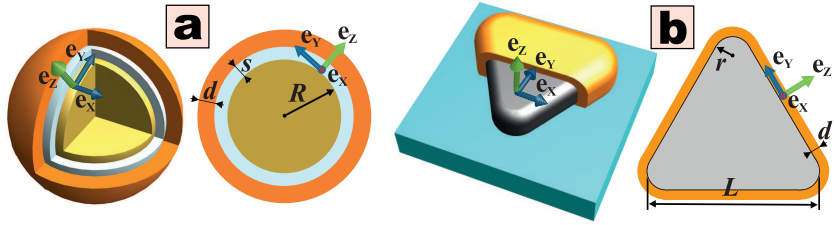}
\end{center}
\caption{Schematic view of a hybrid metal-organic nanostructures under study. Figs.~\ref{fig:shapes}a show a three-layer nanoparticle and Figs.~\ref{fig:shapes}b a two-layer nanoprism. The vectors $\textbf{e}_X$, $\textbf{e}_Y$ and $\textbf{e}_Z $ show the basis for the permittivity tensor of the shell material: $\textbf{e}_Z $ is directed normal to the surface, while $\textbf{e}_X$ and $\textbf{e}_Y$ lie in the tangential plane.}
\end{figure}
\label{fig:shapes}

\section{Theoretical approach}\label{Sec-methods}

The optical spectra of plexcitonic two-layer and three-layer nanoparticles were described in this work using two theoretical methods. We employ the Mie theory extended to the case of multilayer concentric spheres with anisotropic optical properties for spherical nanoparticles. The generalization for a multilayer sphere was made in several works (see, e.g.\cite{Sinzig1994}. To include into consideration the effects of radial anisotropy of the outer organic shell, we use the formulas of the extended Mie theory presented in~\cite{Joannopoulos2010}. For metal-organic nanoparticles of complex shapes (e.g., triangular prisms), our calculations of the extinction cross sections are based on the FDTD method implemented in a freely available software package MIT Electromagnetic Equation Propagation (MEEP). The final results are presented for orientationally averaged cross sections corresponding to the naturally polarized light incident onto the nanoparticles randomly oriented in colloidal solutions or matrices. If the nanostructures are placed on glass we employ an effective medium approximation to simulate the effect of the substrate~\cite{Niu2020, Kond-Leb_OE2019}. The effective permittivity used is $\varepsilon_{\text{h}} = 1.65$. The approach allows us to calculate the cross sections for nanoparticles of arbitrary shapes and dimensions, including contributions from all electric and magnetic multipoles.

Throughout this work, we describe the nanoparticle metallic core by a local and isotropic frequency-dependent dielectric function, $\varepsilon_{\text{m}}(\omega)$. Such an approach is justified since we do not consider here the case of very small nanoparticle sizes, where the effects of spatial dispersion are manifested \cite{Agarwal1983}. The choice of a specific way of describing the dielectric function, $\varepsilon_{\text{m}}(\omega)$, was done similar to our previous works with the account of both contributions from the intraband and interband transitions, $\varepsilon_{\text{m}} \left(\omega \right) = \varepsilon_{\text{intra}} \left(\omega \right) + \varepsilon_{\text{inter}} \left(\omega \right)$, and size effect of Kreibig \cite{Kreibig1995}, associated with scattering of free electrons from an interface between a metallic core and an organic layer.

We note that in all known calculations of the optical properties of hybrid metal/J-aggregate nanoparticles, carried out within the framework of a single exciton purely isotropic models, the frequency-dependent dielectric function of molecular J-aggregates was taken in the scalar form
\begin{equation}
\varepsilon _{\text{J}}(\omega)=\varepsilon^{\infty}_{\text{J}}+\frac{f_{\text{J}} \omega_{\text{J}}^2}{\omega _{\text{J}}^{2}-\omega ^{2}-i\omega {\gamma_{\text{J}}}}.\label{eps-J}
\end{equation}
\noindent
Here $\omega_{\text{J}}$ are the central frequency of the J-band; $\gamma_{\text{J}}$ is its full width; $f_{\text{J}}$ is the reduced oscillator strength; and $\varepsilon^{\infty}_{\text{J}}$ is the permittivity outside the J-band. 

In order to include the anisotropic and orientational effects of the outer J-aggregate shell in the theoretical description of the optical properties of hybrid metal/J-aggregate and metal/spacer/J-aggregate nanoparticles, its local frequency-dependent dielectric function should be significantly modified compared to simplified expression (\ref{eps-J}). In this case, it is necessary to represent it in tensor form and take into account the more complex nature of the J-band, in particular, the presence of several resonance peaks. Accordingly, we assume axial symmetry of the optical properties of the molecular J-aggregate and describe the components of the dielectric function tensor of the organic shell as the sum of the Lorentzians
\begin{equation}
\varepsilon_{(\parallel,\perp)}\left( \omega \right) =\varepsilon^{\infty}_{(\parallel,\perp)} +\sum\limits_{n}\frac{f^{(\parallel,\perp)}_{n}(\omega^{(\parallel,\perp)}_{n})^{2}}{(\omega^{(\parallel, \perp)}_{n})^{2}-\omega^{2}-i\omega \gamma^{(\parallel, \perp)}_{n}}. \label{eps-J_tensor}
\end{equation}
Here $\varepsilon_{\parallel}$ and $\varepsilon_{\perp}$ are the  tensor components associated with the direction parallel to the aggregate orientation axis (\textit{longitudinal} component) and perpendicular to it (\textit{transverse} component), respectively. Parameters in equation (\ref{eps-J_tensor}) are fitted from the measured absorption spectra. 

The fitting procedure is as follows. The first term in the right-hand side of equation (\ref{eps-J_tensor}) is assumed to be the same for $\varepsilon^{\infty}_{\parallel}$, and $\varepsilon^{\infty}_{\perp}$, components of the J-aggregate permittivity tensor. This value is taken from the experimental data. The absorption spectra are simulated using the well-known formula for the dielectric constant of mixtures~\cite{Reynolds1957}. This allows us to evaluate frequencies, $\omega^{(\parallel, \perp)}_{n}$, and widths, $\gamma^{(\parallel, \perp)}_{n}$, of spectral bands as well as the reduced oscillator strengths $f^{(\parallel,\perp)}_{n}$. If the absorption spectra are measured separately for the light polarization parallel and perpendicular to the aggregate orientation axis, the fitting is performed separately for the parameters of $\varepsilon_{\parallel}$ and $\varepsilon_{\perp}$. A more complicated approach is used when the absorption spectra are measured for naturally polarized light or randomly oriented aggregates in a sample. In this case, we consider that some absorption bands are attributed to the longitudinal component and the rest to the transverse component of the J-aggregate permittivity tensor. The absorption spectra, $\sigma_{\parallel}$, and $\sigma_{\perp}$, are computed separately for light polarization parallel and perpendicular to the aggregate axis, respectively. The resulting absorption spectrum is obtained as an equiprobable averaging over the orientations: $\left\langle \sigma\right\rangle = \sigma_{\parallel}/3 + 2\sigma_{\perp}/3$.

In order to explain the available experimental data most completely and correctly, we consider the following cases of molecular arrangements in the J-aggregate shell: (I) \textit{Normal} aggregate orientation, when the direction of $\varepsilon_{\parallel}$ component is parallel to the normal direction to a nanoparticle surface; (II) \textit{Tangential} aggregate orientation, in which the direction of $\varepsilon_{\parallel}$ component is perpendicular to the normal direction, and we assume the equiprobable orientation of the aggregate axis along the tangential plane of a nanoparticle; (III) \textit{Equiprobable} orientation of the aggregate axis in space. In these three cases, the dielectric functions are described by tensors. The corresponding expressions are obtained by performing a proper averaging over possible directions of the J-aggregate axis. With tensor oriented along the $\mathbf{e}_{x}$, $\mathbf{e}_{y}$ and $\mathbf{e}_{z}$ axes (see Fig.~\ref{fig:shapes}) the resulting expressions take the form
\begin{equation}
\resizebox{.93\hsize}{!}
{\begingroup $
\setlength\arraycolsep{0pt}
\widehat{\varepsilon }_{\text{norm}}=\left(
\begin{array}{ccc}
\varepsilon _{\bot } & 0 & 0 \\ 
0 & \varepsilon _{\bot } & 0 \\ 
0 & 0 & \varepsilon _{\Vert } \end{array}
\right) ,\text{\ \ }\widehat{\varepsilon }_{\text{tang}}=\left( 
\begin{array}{ccc}
\frac{1}{2}\varepsilon _{\Vert }+\frac{1}{2}\varepsilon _{\bot } & 0 & 0 \\ 
0 & \frac{1}{2}\varepsilon _{\Vert }+\frac{1}{2}\varepsilon _{\bot } & 0 \\ 
0 & 0 & \varepsilon _{\bot }
\end{array}
\right) ,\text{\ \ }\widehat{\varepsilon }_{\text{epo}}=\left( 
\begin{array}{ccc}
\frac{1}{3}\varepsilon _{\Vert }+\frac{2}{3}\varepsilon _{\bot } & 0 & 0 \\ 
0 & \frac{1}{3}\varepsilon _{\Vert }+\frac{2}{3}\varepsilon _{\bot } & 0 \\ 
0 & 0 & \frac{1}{3}\varepsilon _{\Vert }+\frac{2}{3}\varepsilon _{\bot }
\end{array}
\right). $
\endgroup}
\end{equation} \label{eq:3tensors}
\noindent
In addition, for comparison of our results with those obtained on the basis of standard theoretical models, we consider one more case (IV) that corresponds to purely \textit{isotropic} J-aggregate shell, described by scalar expression~(\ref{eps-J}) for single excitonic peak, or its direct generalization,
\begin{equation}
\varepsilon_\text{iso}\left( \omega \right) =\varepsilon^{\infty}_{\text{J}} +\sum\limits_{n}\frac{f_{n}\omega_{n}^{2}}{\omega_{n}^{2}-\omega^{2}-i\omega \gamma_{n}.}
\label{eps-iso}
\end{equation}
\noindent taking into account the multiple absorption bands of some molecular aggregates.

\section{Results and discussions}

\begin{figure}[htbp]
\centering\includegraphics[width=11.0cm]{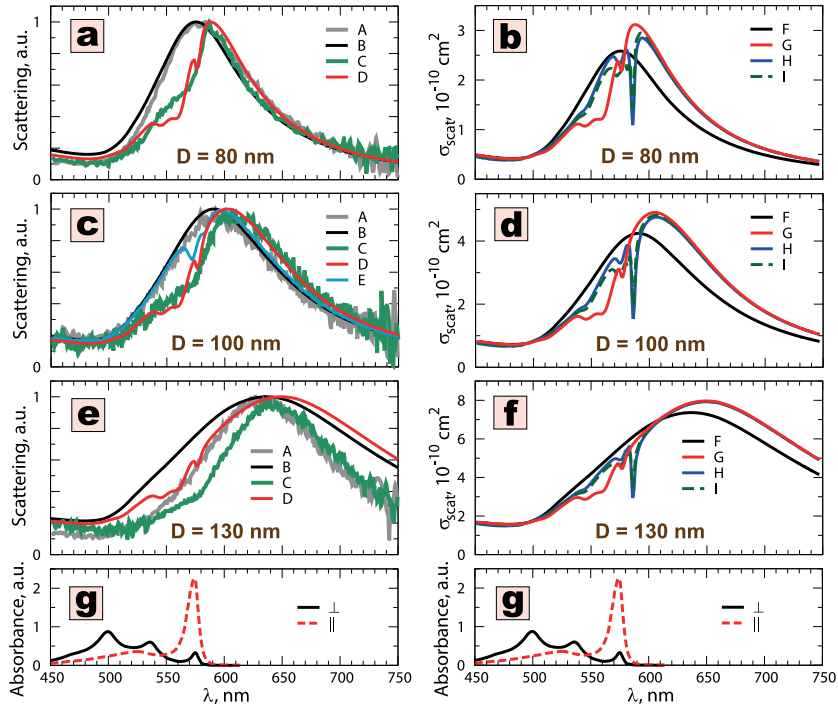}
\caption{Light scattering spectra of spherical three-layer Au/MUA/PIC nanoparticles. Figures \ref{fig:Uwada}a, \ref{fig:Uwada}c, and \ref{fig:Uwada}e in left column show comparisons of our theoretical results with experimental data~\cite{Uwada2007}. Figures \ref{fig:Uwada}b, \ref{fig:Uwada}d, and \ref{fig:Uwada}f in right column show the scattering cross sections calculated for three cases (I) -- (III) of molecular arangement in the J-aggregate shell described in Sec.~\ref{Sec-methods}. The Figures in the top three rows correspond to different diameters of a golden core, $D = 2R$ (see Fig.~\ref{fig:shapes}): $D=80$ nm for \ref{fig:Uwada}a -- \ref{fig:Uwada}b, $D=100$ nm for \ref{fig:Uwada}c -- \ref{fig:Uwada}d, and  $D=130$ nm for \ref{fig:Uwada}e -- \ref{fig:Uwada}f. Curves to show the scattering spectra: Grey curves (A) and black curves (B) are experimental data \cite{Uwada2007} and the results of our calculations for two-layer, Au/MUA, nanoparticles, respectively. Green curves (C) and red curves (D) are experimental data \cite{Uwada2007} and our theoretical results for three-layer Au/MUA/PIC nanoparticles, calculated for \textit{normal} J-aggregate orientation. Cyan curve (E) in Fig.~\ref{fig:Uwada}c show the results of calculations~\cite{Uwada2007} for Au/MUA/PIC nanoparticles obtained using standard scalar J-aggregate dielectric function (\ref{eps-J}). Curves to show the scattering cross sections: black curves (F) -- no outer shell; red curves (G) -- normal J-aggregate orientation; blue curves (H) -- tangential orientation; green dashed curves (I) -- equiprobable orientation of the aggregate in space. Two identical Figures~\ref{fig:Uwada}g show the absorption spectra \cite{Misawa1994} with parallel (dashed red curve) and perpendicular (solid black curve) polarization of incident light relative to the PIC-dye J-aggregate axis.}
\label{fig:Uwada}
\end{figure}

We begin our consideration of the main features in optical spectra of hybrid metal-organic nanostructures associated with the anisotropic properties of their J-aggregate shell with a theoretical interpretation of available experimental data~\cite{Uwada2007} on light scattering spectra by a three-layer, Au/MUA/PIC, nanoparticles (see Figure~\ref{fig:Uwada}). Such a plexcitonic system consists of a spherical gold core, an optically passive self-assembled MUA monolayer, and a J-aggregated outer shell of 1,1`-diethyl-2,2`-cyanine (PIC) dye. The anisotropic optical properties of PIC J-aggregate were studied in~\cite{Scherer1984, Misawa1994, Tani2007, Tani2012, Haverkort2014}. For such oriented J-aggregates, a significant difference was demonstrated in the absorption spectra obtained for parallel and perpendicular polarization of light relative to the aggregate orientation axis (see red and black curves in bottom panels of Fig.~\ref{fig:Uwada}g). To use these experimental data in our theoretical studies, we employ the basic tensor expressions (\ref{eps-J_tensor}) with the parameters fitted from the absorption spectra measured in~\cite{Misawa1994}. The fitting procedure is described in Sec.~\ref{Sec-methods}. Following \cite{Uwada2007} we assume $\varepsilon_{\infty} = 2.9$. Remaining parameters are listed in Table~\ref{tab:pic}.

\begin{table}
\centering
\caption{Parameters of tensor dielectric function of PIC dye J-aggregate.}
\label{tab:pic}
\addtolength{\tabcolsep}{-2pt}
\begin{tabular}{|c|c|c|c|c|}
\hline
$n$ & 1 & 2 & 3 & 4 \\ \hline 
$\omega _{n}^{(\Vert )}$, eV & 2.12 & 2.14 & 2.34 & 2.47 \\ \hline 
$\gamma _{n}^{(\Vert )}$, eV & 0.0077 & 0.075 & 0.23 & 0.25 \\ \hline
$f_{n}^{(\Vert )}$ & 0.24 & 0.083 & 0.11 & 0.046 \\ \hline
\end{tabular}
\ \ \ \ \
\begin{tabular}{|c|c|c|c|c|c|}
\hline
$n$ & 1 & 2 & 3 & 4 & 5 \\ \hline
$\omega _{n}^{(\bot )}$, eV & 2.15 & 2.28 & 2.35 & 2.44 & 2.59 \\ \hline
$\gamma _{n}^{(\bot )}$, eV & 0.017 & 0.063 & 0.21 & 0.099 & 0.29 \\ \hline
$f_{n}^{(\bot )}$ & 0.044 & 0.17 & 0.14 & 0.12 & 0.068 \\ \hline
\end{tabular}
\addtolength{\tabcolsep}{2pt}
\end{table}

Figures \ref{fig:Uwada}a, \ref{fig:Uwada}c, and \ref{fig:Uwada}e (left column) present the results of our calculations of light scattering spectra by Au/MUA/PIC nanoparticles for three different diameters of a gold core ($D = 80$ nm, 100 nm, and 130 nm) and their comparison with experimental data~\cite{Uwada2007}. In our calculations, the thickness of the MUA spacer layer is $s = 1.63$ nm for all core diameters. The thickness of the J-aggregate shell is assumed to be $d = 2$, 2.3, and 2.7 nm for the particles with core diameters 80 nm, 100 nm, and 130 nm, respectively (see Fig.~\ref{fig:shapes}). The  permittivities of MUA and host solution are $\varepsilon_{\text{MUA}} = 2.1$, and $\varepsilon_{\text{h}} = 1.96$, respectively. The theoretical curves in these figures have been obtained using the anisotropic Mie theory for case (I) of \textit{normal} J-aggregate orientation. In addition, in Figs.~\ref{fig:Uwada}b, \ref{fig:Uwada}d, and \ref{fig:Uwada}f (right column), we demonstrate a comparison of the scattering cross sections, $\sigma_{\text{scat}}$, of light from the same nanoparticles, Au/MUA/PIC, calculated for the normal (I), tangential (II) and equiprobable (III) orientations of the outer J-aggregate shell of PIC-dye relative to the surface of a gold core covered with MUA spacer layer. As is evident from the figures, the shapes of curves for these three cases turn out to be quite different, and only case (I) of normal J-aggregate orientation agrees well with the experimental data \cite{Uwada2007}. These differences in the shape of the theoretical curves are, in fact, a consequence of radical differences in character and shapes of the absorption spectra obtained for parallel and perpendicular polarization of PIC-dye J-aggregate axis (see red and black curves in the bottom of Fig.~\ref{fig:Uwada}g). Here it is important to stress that if the scattering spectra were calculated without taking into account the anisotropic properties of the J-aggregate, the narrow resonance peak in permittivity, $\varepsilon_{\parallel}$, would appear as a specific narrow dip in the scattering spectra, as directly shown by curve E in Fig.~\ref{fig:Uwada}c.

\begin{figure}[htbp]
\centering\includegraphics[width=11.0cm]{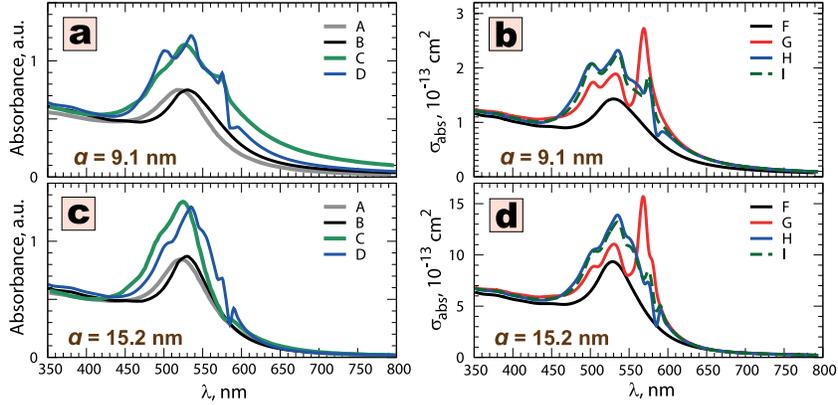}
\caption{Spectra of light absorption by Au/TMA/PIC nanoparticles. Figures~\ref{fig:Yoshida}a and \ref{fig:Yoshida}c (left column) show a comparison of our theoretical results with experimental data from Supporting Information of~\cite{Yoshida2009a}. Figures~\ref{fig:Yoshida}b and \ref{fig:Yoshida}d (right column) show the absorption cross sections calculated for three cases (I) -- (III) of molecular arrangement in the J-aggregate shell. The results are presented for the average diameters of Au/TMA particles $a = 9.1$ nm (Figs.~\ref{fig:Yoshida}a -- \ref{fig:Yoshida}b), and $a = 15.2$ nm (Figs.~\ref{fig:Yoshida}c -- \ref{fig:Yoshida}d), as estimated in~\cite{Yoshida2009a}. In Figs.~\ref{fig:Yoshida}a and \ref{fig:Yoshida}c (left column): Grey curves (A) and black curves (B) are experimental data and the results of our calculations for Au/TMA, respectively. Green curves (C) and blue curves (D) are experimental data and our theoretical results for Au/TMA/PIC, calculated for \textit{tangential} J-aggregate orientation. Notations of curves in Figs.~\ref{fig:Yoshida}b and \ref{fig:Yoshida}d (right column) are same as in Figs.~\ref{fig:Uwada}b, \ref{fig:Uwada}d, and \ref{fig:Uwada}f.}
\label{fig:Yoshida}
\end{figure}

Next, we calculate the light absorption spectra of metal-organic three-layer nanoparticles, Au/TMA/PIC. This system is experimentally studied in~\cite{Yoshida2009a}. It differs essentially from the example considered above by the material of the spacer layer, which is TMA: N,N,N-trimethyl(11-mercaptoundecyl) ammonium chloride. Another difference is an order of magnitude smaller size of the particle core. In our calculations, we use the following parameters: for the average particle diameter of Au/TMA, $a = 9.1$ nm, we take $D = 2R = 7.1$ nm, $s = 1.0$ nm, and $d = 1.0$ nm; while for the average particle diameter, $a = 15.2$ nm, we take $D = 2R = 12.4$ nm, $s = 1.4$ nm, and $d = 1.2$ nm (see Fig.~\ref{fig:shapes}). The permittivities of TMA and host solution are $\varepsilon_{\text{TMA}} = 2.4$, and $\varepsilon_{h} = 1.78$, respectively. The results for Au/TMA/PIC are shown in Fig.~\ref{fig:Yoshida}. Similar to Fig.~\ref{fig:Uwada} panels in the left column of Fig.~\ref{fig:Yoshida} present a comparison of our theoretical results with experimental data from Supporting Information of~\cite{Yoshida2009a}. The panels in the right column demonstrate the behavior of the absorption cross sections calculated for three different cases (I--III) of the molecular J-aggregate orientation in the outer shell of the composite nanoparticle.

For Au/TMA/PIC nanoparticles, the agreement with the experimental data is obtained for the case of \textit{tangential} J-aggregate orientation in the organic shell. Whereas, for Au/MUA/PIC, \textit{normal} J-aggregate orientation yields better agreement. The difference in aggregate orientation of the organic shell can be explained by a significant difference between the chemical properties of MUA and TMA surfaces, on which the PIC J-aggregate is assembled. It is known that coating the Au or Ag nanoparticles with MUA and TMA produces the particles of opposite charge~\cite{Kalsin2006, Kowalczyk2021}. When the gold nanoparticle is coated with the TMA layer, the positively charged particle is formed, while the coating of the same particle with the MUA layer forms a negatively charged one. It was demonstrated~\cite{Kalsin2006} that such charging could cause electrostatic self-assembly of sphalerite (diamond-like) crystals, in which each nanoparticle has four oppositely charged neighbors. It was also shown \cite{Kowalczyk2021} that the oppositely charged nanoparticles with the same metallic core but TMA or MUA coating could function as universal surfactants that control the growth and stability of microcrystals of charged organic molecules. Thus, our conclusion that the J-aggregate of a PIC dye has different arrangements in the outer shell, depending on the material of the spacer layer (TMA or MUA), agrees with concepts discussed in \cite{Kalsin2006, Kowalczyk2021}.

Analysis of Figs.~\ref{fig:Uwada} and \ref{fig:Yoshida} shows that the results calculated for tangential and equiprobable orientations are close to each other, while the results obtained for normal orientation are different. This observation can be explained by considering the form of permittivity tensors. Indeed, X- and Y- components of both $\widehat{\varepsilon }_{\text{tang}}$ and $\widehat{\varepsilon }_{\text{epo}}$ tensors contain terms, which mix the values of $\varepsilon _{\Vert}$ and $\varepsilon _{\bot}$ in different proportions. In turn, the $\widehat{\varepsilon }_{\text{epo}}$ tensor has the same values at all diagonal terms. Thus, the results obtained with this tensor are equivalent to those calculated with some effective scalar isotropic dielectric function (\ref{eps-J}). If the J-aggregate absorption spectra contain one dominant band, its oscillator strength can be adjusted to fit the main spectroscopic features of a hybrid system with tangentially oriented aggregates. This clearly explains why models with isotropic shells could be useful for treating a number of experimental data for spherical nanoparticles, except for some specific cases, like Au/MUA/PIC system~\cite{Uwada2007}.

The use of an anisotropic shell model allows us to treat the recent experimental data \cite{Takeshima2020} on the optical properties of hybrid AgPR/TPP nanoplatelets consisting of silver nanoprisms with different aspect ratios coated with TPP dye (5,10,15,20-Tetraphenyl-21H,23H-porphyrin). The authors of work~\cite{Takeshima2020} developed a unique technique and fabricated the triangular silver nanoprisms with precisely tuned localized surface plasmon resonance wavelengths. This new technique has a significant advantage over some previous works on plasmonic and plexcitonic nanoprisms because of the relatively small dispersion in size of the produced nanoplatelets. This achievement allows studying the fine details in the plexcitonic spectra by avoiding inhomogeneous broadening. The TPP dye has four absorption peaks in the wavelength range of 500–700 nm so that the hybrid AgPR/TPP system exhibits multi-mode plasmon-exciton coupling.

\begin{figure}[htbp]
\centering\includegraphics[width=11.0cm]{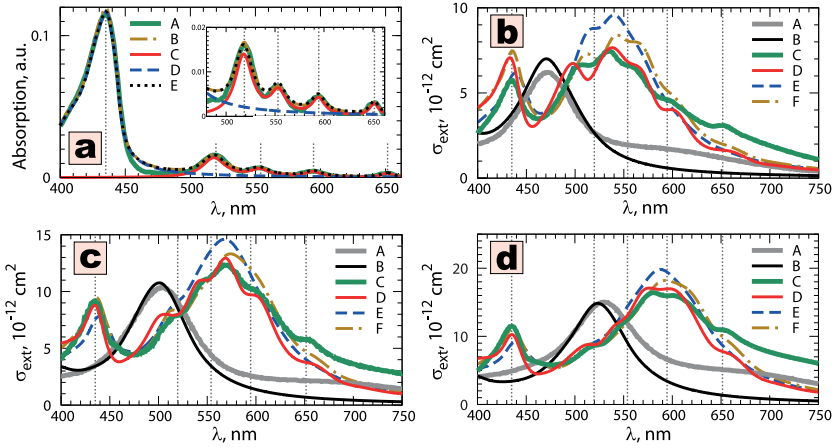}
\caption{Results of simulation of the experimental data~\cite{Takeshima2020}. Figure~\ref{fig:AgPRs-TPP}a shows the absorption spectra of TPP: Green curve (A) -- experimental data~\cite{Takeshima2020}. Yellow dashed-dotted curve (B) -- results of fitting with the isotropic expression (\ref{eps-iso}). Red curves (C) -- the contribution of the longitudinal component of the model anisotropic tensor, $\sigma_{\parallel}/3$. Blue dashed curve (D) -- the contribution of the transverse component $2\sigma_{\perp}/3$. Black dotted curve (E) -- resulting spectra $\left\langle \sigma\right\rangle$. Figures~\ref{fig:AgPRs-TPP}b -- ~\ref{fig:AgPRs-TPP}d show light extinction cross sections by hybrid Ag/TPP nanoprisms computed with anisotropic TPP shell model. The nanoprism sizes correspond to those denoted in~\cite{Takeshima2020} as AgPRs-500, AgPRs-540 and AgPRs-560, respectively. In each figure: Grey curves (A) and black curves (B) show scaled experimental and orientally averaged theoretical results for bare AgPRs, respectively. Results for immobilized hybrid AgPR/TPP: Green curves (C) -- scaled experimental data. Red curves (D) -- computed spectra for \textit{normal} aggregate orientation. Blue dashed curves (E) -- results obtained for \textit{tangential} orientation. Yellow dashed-dotted curves (F) -- results obtained for \textit{isotropic} outer shell material. In each figure vertical grey dotted lines mark absorption wavelengths of TPP.}
\label{fig:AgPRs-TPP}
\end{figure}

We succeeded in the theoretical treatment of the complicated optical spectra of AgPR/TPP nanoplatelets by considering the separation of the absorption spectra of the TPP dye into a sum of two contributions from longitudinal and transverse permittivity components as shown in Fig.~\ref{fig:AgPRs-TPP}a. The high absorption band around $\lambda=434$ nm is attributed to the transverse component of the dielectric tensor, $\varepsilon_{\perp}\left( \omega \right)$, while the four smaller bands to the longitudinal component, $\varepsilon_{\parallel}\left( \omega \right)$. The assumption that the transverse components of the dielectric tensor cause more intense absorption, then the longitudinal component agrees with the symmetric flat molecular structure of TPP, reported in~\cite{Takeshima2020}. We also consider the purely isotropic case with the scalar dielectric function of the TPP dye described by expression (\ref{eps-iso}).

We obtain the parameters of the tensor components in (\ref{eps-J_tensor}) and scalar (\ref{eps-iso}) by fitting the TPP absorption spectra using the approach described in \ref{Sec-methods}. The constant $\varepsilon_{\infty}$ is taken to be 2.56~\cite{Takeshima2020}. The rest values found are presented in Table~\ref{tab:tpp}.

\begin{table}
\centering
\caption{Parameters of tensor dielectric function of TPP dye.}
\label{tab:tpp}
\addtolength{\tabcolsep}{-2pt}
\begin{tabular}{|c|c|c|c|c|}
\hline
$n$ & 1 & 2 & 3 & 4 \\ \hline 
$\omega _{n}^{(\Vert )}$, eV & 2.39 & 2.24 & 2.09 & 1.90 \\ \hline 
$\gamma _{n}^{(\Vert )}$, eV & 0.098 & 0.074 & 0.054 & 0.031 \\ \hline
$f_{n}^{(\Vert )}$ & 0.037 & 0.013 & 0.0083 & 0.0042 \\ \hline
$f_{n}^{\text{iso}}$ & 0.025 & 0.0065 & 0.0041 & 0.0021 \\ \hline
\end{tabular}
\ \ \ \ \
\begin{tabular}{|c|c|c|c|c|c|}
\hline
$n$ & 1 & 2 & 3 & 4 \\ \hline
$\omega _{n}^{(\bot )}$, eV & 2.79 & 2.82 & 2.88 & 3.04 \\ \hline
$\gamma _{n}^{(\bot )}$, eV & 0.045 & 0.090 & 0.17 & 0.29 \\ \hline
$f_{n}^{(\bot )}$ & 0.034 & 0.077 & 0.063 & 0.053 \\ \hline
$f_{n}^{\text{iso}}$ & 0.034 & 0.077 & 0.063 & 0.053 \\ \hline
\end{tabular}
\addtolength{\tabcolsep}{2pt}
\end{table}

Figures~\ref{fig:AgPRs-TPP}b -- \ref{fig:AgPRs-TPP}d show a comparison of experimental data \cite{Takeshima2020} with the results of our calculations. The results are presented for the nanoprism sizes, denoted in~\cite{Takeshima2020} as AgPRs-500 (Fig.~\ref{fig:AgPRs-TPP}b), AgPRs-540 (Fig.~\ref{fig:AgPRs-TPP}c) and AgPRs-560 (Fig.~\ref{fig:AgPRs-TPP}d), for which strong plasmon-exciton coupling is achieved. The nanoprism sizes used in our calculations are as follows: for AgPRs-500 -- $L = 22.5$ nm, for AgPRs-540 -- $L = 27$ nm, and for AgPRs-560 -- $L = 31$ nm (see Fig.~\ref{fig:shapes}b). Corner radii and silver core thickness are same for all prisms, $r = 4$ nm and $h = 10$ nm, respectively. To mimic the experimental conditions, we considered nanoprisms immobilized on the glass surface so that the organic shell covered only the top and sides of a silver core (see Fig.~\ref{fig:shapes}b). The thickness of the TPP shell is $d = 7$ nm for all samples, which corresponds to the height of the scattered molecular aggregates measured in~\cite{Takeshima2020}. Thus, all the sizes used in our simulations agree with those reported in~\cite {Takeshima2020}. We also evaluated the extinction cross sections for the case of nanoprisms, which are uniformly covered with the organic shell on all sides with shell thickness $d = 4$ nm. The results obtained for this case are similar to those obtained for the configuration mimicking the experimental setup. To convert from arbitrary units to centimeters squared we multiply the experimental curves in Figs.~\ref{fig:AgPRs-TPP}b -- ~\ref{fig:AgPRs-TPP}d by factors: $3.2\cdot 10^{-11}$ cm$^{2}$, $5.2\cdot 10^{-11}$ cm$^{2}$ and $7.3\cdot 10^{-11}$ cm$^{2}$, respectively.

Considering the organic shell anisotropic properties, a much better agreement between the theoretical results and the experimental data is achieved. In particular, by assuming normal aggregate orientation, multiple peaks and dips are correctly reproduced in the range of 500–700 nm. Such agreement can not be achieved within the purely isotropic model of molecular aggregate shell because the ratios of the reduced oscillator strengths are fixed according to the data on the TPP absorption. Overestimation of the reduced oscillator strengths causes an increase in frequency difference between the intense left peak ($\sim435$ nm) and the main broad absorption band at the longer wavelengths. Overestimation of the shell thickness causes a shift of the main absorption band towards the longer wavelengths because of the relatively high value of $\varepsilon_{\infty} = 2.56$~\cite{Takeshima2020}. It is worth mentioning that the normal orientation of the molecular aggregate corresponds to the case when a dominating dipole moment component is parallel to the nanostructure surface. Assuming that a larger dipole component corresponds to larger aggregate dimensions, we can conclude that the molecular aggregate is flat, and the plane of the aggregate is laid flat on the surface. As is mentioned above, this hypothesis agrees with the molecular structure of TPP dye, reported in~\cite{Takeshima2020}.

Below we study the main spectroscopic features caused by shell anisotropy. First, we consider a core-shell spherical nanoparticle with a silica core ($\varepsilon_{\text{c}} = 2.4$) and a J-aggregate shell with a single Lorentzian spectral band (\ref{eps-J}) in the longitudual component, $\varepsilon_{\parallel}$ and $\varepsilon_{\perp} = \varepsilon^{\infty}_{\text{J}}$. We take the parameters of J-aggregate to be those of TDBC dye~\cite{Bellessa2009}: $\varepsilon^{\infty}_{\text{J}}$ = 2.56, $\hbar \omega_{\text{J}}=2.11$ eV ($\lambda_{\text{J}}=587.6$ nm), $\gamma_{\text{J}}=0.03$ eV, and $f_{\text{J}}=0.41$. The host medium is taken to be water, $\varepsilon_{\text{h}} = 1.78$.

\begin{figure}[htbp]
\centering\includegraphics[width=11.0cm]{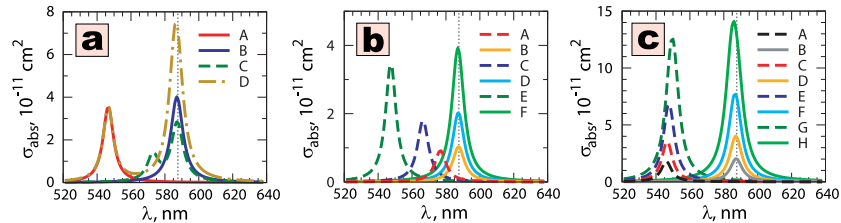}
\caption{Absorption spectra by SiO$_{2}$/J-aggregate nanoparticle calculated for different J-aggregate orientations. Figure~\ref{fig:SiO2-TDBC}a -- results for the parameters of the TDBC J-aggregate. Red curve (A) -- normal aggregate orientation. Blue curve (B) -- tangential orientation. Green dashed curve (C) -- equiprobable orientation. Yellow dashed-dotted curve (D) -- isotropic shell. Figure~\ref{fig:SiO2-TDBC}b -- results for different oscillator strengths, $f_{\text{J}}$. Dashed curves are for normal and solid curves are for tangential orientations: $f_{\text{J}} = 0.1$ -- dashed red (A) and solid orange (B) curves; $f_{\text{J}} = 0.2$ -- dashed blue (C) and solid cyan (D) curves; $f_{\text{J}} = 0.4$ -- dashed dark-green (E) and solid green (F) curves. Figure~\ref{fig:SiO2-TDBC}c -- results for different shell thickness, $d$. Dashed curves are for normal and solid curves are for tangential orientations: $d = 1$ nm -- dashed black (A) and solid grey (B) curves; $d = 2$ nm -- dashed red (C) and solid orange (D) curves; $d = 4$ nm -- dashed blue (E) and solid cyan (F) curves; $d = 8$ nm -- dashed dark-green (G) and solid green (H) curves. Vertical grey dotted line marks the TDBC J-aggregate absorption wavelength, $\lambda_{\text{J}} =587.6$ nm.}
\label{fig:SiO2-TDBC}
\end{figure}

Figure~\ref{fig:SiO2-TDBC}a shows absorption spectra of such model nanoparticle for four cases (I)--(IV) of molecular arrangements in the J-aggregate shell described in Sec.~\ref{Sec-methods}. The particle core diameter is 100 nm, shell thickness is 2 nm. One can mention that absorption wavelengths differ substantially for normal and tangential orientations. In cases of equiprobable orientation and isotropic shell, spectra have two bands associated with both normal and tangential components. In the model under study, the permittivity tensors for cases of equiprobable orientation and isotropic shell differ only by values of reduced oscillator strengths (see (\ref{eq:3tensors})--(\ref{eps-iso})). For equiprobable orientation, this value is three times smaller than for the isotropic shell. Thus, a comparison of the curves for these cases shows that the value of the oscillator strength significantly affects the absorption wavelengths. This conclusion is confirmed by comparing absorption spectra evaluated for the same sizes and composition of nanoparticles, but for various values of J-aggregate oscillator strength, shown in Fig.~\ref{fig:SiO2-TDBC}b. For normal aggregate orientation, the absorption wavelength depends linearly on the oscillator strength, while for tangential orientation, the wavelength remains close to the J-aggregate absorption wavelength. The results shown in Fig.~\ref{fig:SiO2-TDBC}c demonstrate that the thickness of the shell does not significantly affect the absorption wavelengths.

Analyzing Figs.~\ref{fig:SiO2-TDBC} we conclude that the absorption spectra of a silica nanoparticle coated with a J-aggregate contain one or two narrow absorption bands depending on the molecular arrangements in the shell. A single excitonic absorption wavelength is observed only for normal and tangential aggregate orientation. The absorption wavelengths of this system strongly depend on both the aggregate orientation and the reduced oscillator strength. In the previous subsection, we concluded that the J-aggregate arrangements in the organic shell of the three-layer hybrid structures strongly depend on the physico-chemical properties of the intermediate optically passive layer (spacer). Combining this conclusion with the above analysis allows us to propose a three-layer plexcitonic nanoparticle, which consists of a dielectric (or semiconductor) core, an optically passive self-assembled layer, and a J-aggregated outer shell as the most convenient system for experimental study of the anisotropic effects of molecular J-aggregate shell on the optical properties of a hybrid system. Such systems have two advantages. First, molecular arrangements in the organic shell could be controlled by changing the material of the intermediate spacer layer. Second, the molecular arrangement could be straightforwardly determined by measuring the absorption wavelengths.

\begin{figure}[htbp]
\centering\includegraphics[width=11.0cm]{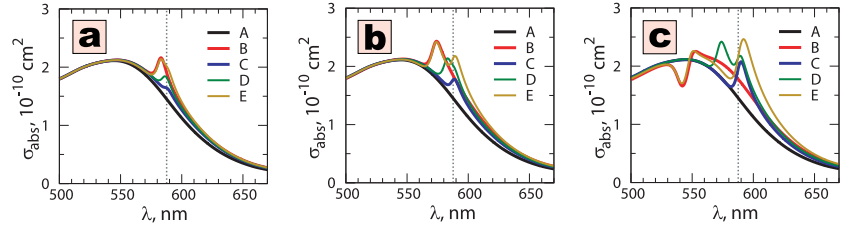}
\caption{Absorption spectra by Au/J-aggregate nanoparticle calculated for different orientations and oscillator strengths of the J-aggregate: Figure~\ref{fig:Au-TDBC}a -- $f_{\text{J}} = 0.05$, Fig.~\ref{fig:Au-TDBC}b --  $f_{\text{J}} = 0.15$, Fig.~\ref{fig:Au-TDBC}c -- $f_{\text{J}} = 0.45$. Black curve (A) -- no shell. Red curve (B) -- normal aggregate orientation. Blue curve (C) -- tangential orientation. Green curve (D) -- equiprobable orientation. Yellow curve (E) -- isotropic shell. The sizes of nanoparticles: $D = 2R = 100$ nm, $d = 2$ nm. Vertical grey dotted line marks the TDBC J-aggregate absorption wavelength, $\lambda_{\text{J}} =587.6$ nm.}
\label{fig:Au-TDBC}
\end{figure}

Figure~\ref{fig:Au-TDBC} shows absorption spectra of the two-layer system, which is similar to that studied above. The sizes and the shell material are the same, but the core material is gold instead of silica.  Similar to Figs.~\ref{fig:SiO2-TDBC}a we present our results for four considered cases of molecular arrangements in the J-aggregate shell. As in the case of silica core, the spectral features specific for different J-aggregate orientations are also observed in the system with the golden core. However, as can be seen in Fig.~\ref{fig:Au-TDBC}c under certain conditions, the absorption band, associated with normal molecular aggregate orientation, could be strongly coupled with the core plasmon, causing the spectral dip formation instead of the peak. At the same time, the absorption band, associated with tangential orientation, remains weakly coupled with the core plasmon. The dependences of the absorption wavelength on the oscillator strength for normal and tangential orientations are similar to those observed for the system with silica core. When the orientation is tangential, the wavelength of spectral peak (or dip) is close to the J-aggregate absorption wavelength, while in the case of normal orientation, the associated peak (or dip) tends to shift towards shorter wavelengths with an increase of the oscillator strength. We note that this observation also could be used to identify the anisotropic shell effects in the optical spectra of hybrid nanoparticles.

\section{Conclusions}

We have reported novel results on the effects of near-field electromagnetic coupling between a localized surface plasmon in the metallic core and a J-aggregate exciton in the dye shell of two-layer and three-layer metal-organic nanoparticles. The key point of our theoretical consideration consists in the self-consistent description of the optical spectra of such hybrid nanoparticles of spherical and complex shapes by taking into account the tensorial form of the local dielectric function of the molecular aggregate shell associated with the anisotropic and orientational effects in the molecular arrangements.

The approach used in the present work allows us to explain some available experimental data on the optical spectra of metal/spacer/J-aggregate nanoparticles and metal/J-aggregate nanoplatelets. We have illustrated this fact with three examples by performing detailed numerical calculations of the absorption and scattering spectra of three-layer nanospheres (Au/MUA/PIC and Au/TMA/PIC) and two-layer nanoprisms (Ag/TPP). Similar calculations and theoretical analysis can be made for a variety of plexcitonic multilayer metal-organic nanosystems possessing anisotropic dielectric properties of their J-aggregated excitonic shell.

The main result of this work is that there are cases when the anisotropy of the permittivity of the excitonic shell drastically affects the light absorption and scattering spectra of hybrid plexcitonic nanoparticles. The effect manifests itself in qualitatively different shapes of these spectra for the cases of normal and tangential orientations of aggregated molecules in the dye shell relative to the nanostructure surface. It is shown that the optical spectra of a hybrid nanoparticle obtained with only one distinct aggregate orientation yield good agreement with the available experimental data.

We have demonstrated a strong dependence of the molecular aggregate orientation of a hybrid nanoparticle on the surface material on which the aggregate is assembled. It was illustrated by comparing our calculations with the experimental data on the optical spectra of Au/MUA/PIC and Au/TMA/PIC nanospheres having quite different MUA and TMA spacer layers. The difference in aggregate orientation can be explained by significantly different chemical properties of MUA and TMA surfaces. This conclusion agrees with the current concepts in chemistry that the metallic nanoparticles coated with MUA and TMA form oppositely charged particles. Our results indicate that it is possible to effectively control the optical properties of hybrid nanostructures by correctly selecting the optically passive spacer layer. It is worthwhile to point out that although the optical properties of the spacer insignificantly affect the resulting spectra of the hybrid system, its material affects the anisotropic properties of the outer shell. This, in turn, strongly affects the spectra.

We have clarified when and why the use of the isotropic scalar dielectric function (\ref{eps-J}) of the outer organic shell can provide a reasonable explanation for some available experimental data. We have shown that the results calculated for tangential and equiprobable orientations are close to each other in some cases, while the results obtained for normal orientation are different. Thus, if the J-aggregate absorption spectra possess only one excitonic peak, its reduced oscillator strength can be adjusted to reproduce the basic spectroscopic properties of a hybrid system with tangentially oriented aggregates. This clearly explains why models with isotropic shells can be useful for describing a number of experimental data for spherical metal-organic nanostructures, except for some specific cases, like Au/MUA/PIC nanoparticles and Ag/TPP nanoplatelets.

We have proposed a type of hybrid nanostructures that seems most convenient for observing the anisotropic shell effect on its optical properties. This is a three-layer plexcitonic nanoparticle consisting of a dielectric or a semiconducting core, an optically passive layer, and a J-aggregated outer shell. The spectrum of such a system contains narrow absorption bands, with wavelengths strongly dependent on the aggregate orientation in a shell. When the orientation is tangential, the wavelength of the corresponding spectral band is close to the aggregate absorption wavelength, while for the normal orientation, its spectral peak tends to shift towards a short wavelength range. Thus, the molecular arrangement can be directly determined by measuring the absorption wavelengths. At the same time, the aggregate orientation can be controlled by changing the material of the spacer layer.

We have also shown that in metal-organic structures, the Frenkel excitons corresponding to normal and tangential molecular aggregate orientations separately interact with the localized surface plasmon. In some cases, the exciton associated with some aggregate orientation can be strongly coupled with the plasmon, and these cause the spectral dip formation instead of the peak. At the same time, the exciton associated with the perpendicular orientation remains weakly coupled with the core plasmon.

In summary, the results of the work have demonstrated for the first time vivid manifestations of the anisotropic and orientational effects associated with ordered molecular aggregate shells in optics and spectroscopy of hybrid plexcitonic nanostructures. Understanding the role of these effects is necessary for approaching a more complete and clear picture of physical phenomena occurring as a result of the electromagnetic coupling of Frenkel excitons with localized plasmons in hybrid metal-organic nanoparticles, as well as in their complexes (dimers, trimers, quadrumers, etc.) and arrays. Further studies in this direction are promising for the development of some photonic and optoelectronic devices of the next generation operating on the basis of the effects of near-field electromagnetic plexcitonic coupling.

\section*{Funding} Russian Science Foundation (project No 19-79-30086).

\section*{Disclosures} The authors declare no conflicts of interest.

\section*{Acknowledgments}
We acknowledge funding support listed above. The authors are grateful to A.A. Narits for the valuable discussions.

\end{document}